\documentclass[aps,pra,twocolumn]{revtex4}
\usepackage{psfig}
\def\trace{{\rm Tr}}

\def\hcal{{\cal H}}
\def\bcal{{\cal B}}

\def\duzomniejsze{<\kern-.7mm<}
\def\duzowieksze{>\kern-.7mm>}

\def\textbf#1{{\bf #1}}
\def\beq{\begin{equation}}
\def\eeq{\end{equation}}
\def\be{\begin{equation}}
\def\ee{\end{equation}}
\def\ben{\begin{eqnarray}}
\def\een{\end{eqnarray}}
\def\beqa{\begin{eqnarray}}
\def\eeqa{\end{eqnarray}}
\def\eea{\end{array}}
\def\bea{\begin{array}}
\newcommand{\bei}{\begin{itemize}}
\newcommand{\eei}{\end{itemize}}
\newcommand{\bee}{\begin{enumerate}}
\newcommand{\eee}{\end{enumerate}}

\def\m{{\otimes m}}
\def\<{\langle}
\def\>{\rangle}

\def\dt#1{{{\kern -.0mm\rm d}}#1\,}
\def\ypodpis{\raise4mm\hbox{$\omega$}}
\def\rhon{\varrho^{\otimes n}}
\def\sigman{\sigma^{\otimes n}}

\def\rhonab{\varrho^{\otimes n}_{AB}}
\def\prodm{P_{|00\>}^\m}

\def\rab{\rho_{AB}}

\begin{document}
\title{Compressing compound states}
%\title{Concentration of information to local form and data compression}
\author{Micha\l{} Horodecki$^{(1)}$, Pawe\l{} Horodecki$^{(2)}$ and 
Jonathan Oppenheim$^{(1)(3)}$ }

\affiliation{$^{(1)}$Institute of Theoretical Physics and Astrophysics,
University of Gda\'nsk, Poland}

\affiliation{$^{(2)}$Faculty of Applied Physics and Mathematics,
Technical University of Gda\'nsk, 80--952 Gda\'nsk, Poland}

\affiliation{$^{(3)}$
Racah Institute of Theoretical Physics, 
Hebrew University of Jerusalem, Givat Ram, 
Jerusalem 91904, Israel}

\begin{abstract}
Quantum compression can be thought of not
only as compression of a signal, but also as 
a form of cooling.  In this view, one is
interested not in the signal, but in obtaining
purity. In compound systems, one may be interested to cool the system
to obtain {\it local purity} by use of local operations and 
classical communication [Oppenheim et al. Phys. Rev. 
Lett. {\bf 89},180402 (2002)]. Here we compare it with 
usual compression and find that it can be represented as compression 
with suitably restricted means. 
\end{abstract}
\maketitle

\section{Introduction}

The technique of quantum date compression \cite*{Schumacher1995}
(cf.  \cite*{PetzMosonyi,HiaiPetz91}), can be used to tranform a
signal from a quantum source onto a smaller number of qubits in
order to obtain more efficient storage or transmision.
In this scenario, one is interested in preserving the
signal, and one discards the redundant qubits.  These
redundant qubits are then thought of as containing no
information, and can be regarded as being in some known
(or standard) state.
One might also be interested in applying this technique
in another situation, namely, in the case where one
treats the signal as noise, and instead, one is interested
in obtaining pure states in a standard state.  Such a 
situation was considered in
\cite*{VaziraniS1998} where it was applied to produce 
more pure initial states for a NMR quantum computer.
In this case, ``the signal'' is actually noise and
one compresses and discards this noise to isolate
the pure states (see \cite*{Beth-thermo} in this context). 

Such a technique is also useful from the point of view
of a paradigm introduced in \cite*{OHHH2001}. 
There, one considers parties in distant labs who
share a state $\rab$, and who wish to distill 
local pure states.  This can be thought
of as the complementary procedure \cite*{compl}
to the usual situation, where they attempt to distill
entanglement.  It was found that this paradigm allows
one to understand nature of correlations in shared quantum states.
The insight came from thermodynamics, where pure states
in a known state can be treated as a resource and used
to extract work from a single heat bath\cite*{szilard}.
A more information theoretic analysis was done in \cite*{nlocc}
and \cite*{uniqueinfo} where the techniques of quantum information,
including compression where applied to such a paradigm.

Here, we expand on the ideas introduced in \cite*{nlocc}
and show in greater detail how to apply techniques
developed by Rains \cite*{Rains2001} for entanglement theory, to the
paradigm of distilling pure product states.
This allows to express the problem of distilling local information 
in terms of compression with restricted means. 

%More generally, the techniques presented here
%allows us to look at the compression rate
%when our means are somehow restricted, whether
%because of having to use only local operations and
%classical communication (LOCC) or because of some
%other restriction.

\section{Concentration of subjective and objective information}
\label{sec:single-system}

In this section we will discuss the dual pictures of  
compression of quantum information. We will follow the 
approach of Ref. \cite*{uniqueinfo}.  
Let us first calculate the rate of compression in the general
case where one is free to perform all operations.

Given a state $\rhon$, with $\varrho$ acting on Hilbert space $C^d$ 
one can ask for the smallest Hilbert space 
that still carries most of the weight of the state. Such a Hilbert space 
is called the {\it typical subspace}. In other words we would like to know 
the minimal dimension of the projector $P$ satisfying 
\be
\trace\rhon P\geq 1-\epsilon. 
\label{eq:highfid}
\ee
The interesting characteristic is the number of qubits of the 
typical subspace, per input copy $\varrho$
\be
R_\epsilon = \lim_n {1\over n}\log \trace P.
\ee
where $P$ is chosen to have minimal dimension under the restriction 
(\ref{eq:highfid}).
It turns out \cite*{HiaiPetz91} that 
for $\epsilon\in(0,1)$,  
$R_\epsilon$  does not depend on $\epsilon$ and is equal to 
the von Neumann entropy of $\varrho$
\be
R_\epsilon\equiv R=S(\varrho)
\ee
The rate $R$ can be interpreted as the amount of qubits needed to carry 
the signal 
produced by the source, for which the  density matrix of $n$  
messages is $\rhon$. The actual scheme of transmission of   
the information from the source is as follows. One makes a measurement 
given by the identity resolution $\{P,I-P\}$ and with high probability 
the outcome corresponding to $P$ is obtained.  Then the resulting state 
collapses onto the typical subspace (the subspace of the projector $P$). 
 Thus the support of the state has now dimension $2^{nS}$. 
The total system can be now unitarily transformed into the state 
$\varrho'\otimes|\psi\>\<\psi|$. 
The state $\varrho'$ 
carries the information, while the state $|\psi\>\<\psi|$ 
represents redundancies. The state resides on the tensor product of 
two spaces: the
"signal  space" $\hcal_s$ consisting of $nS$ qubits  and the "redundancy 
space" $\hcal_r$ consisting of $n(\log d - S)$  qubits. 

After we have perform concentration of 
information  we have two opposite schemes. In the 
"subjective" scenario information is randomness,
and  it  is represented  by the signal part. One then keeps this part, 
while discarding redundancies. This is the usual interpretation of
Schumacher compression.  The signal contains the information
because it contains something which we have to read
in order to learn about it.  The more random something is,
the more information it has.  We are {\it surprised} by its
contents.

In the opposite scheme -- the "objective" scenario -- the information 
is {\it purity}. A known pure state contains information because we
know what the state it.  A standard pure state is something 
which is known to us, hence we consider it to be information.
Here, the ``signal part'' is treated as noise, and is rejected, while 
the redundancy part represents a valuable resource  - qubits in 
pure states. The former approach is in spirit of Shannon, while 
the latter one is of thermodynamic origin and is in the spirit of 
Brillouin and Szilard.

Note that in the first scenario, one has to be careful that 
the operation does not damage the information of the signal part.
Thus the compression is  attained  by a highly degenerated measurement,
so that the pure states - individual signals - will not 
loose their quantum coherences.  However  if we are interested 
in the pure part, then we can make a measurement  that is non-degenerate.
In particular it can be a complete measurement, with one dimensional 
projectors.  Then, if the noise  were  actually some 
"quantum information",
it would get totally destroyed. But the pure part would remain 
untouched (if the measurement is chosen in such a way that 
$|\psi\>$ corresponds to possible outcomes).

In this paper we are interested in the "objective" scenario, 
and information for us is "purity". The rate of obtaining 
pure qubits is given by 
\be
R^{conc}=\log d - S=\log d - \lim_n {1\over n} \log \trace P\rhon
\ee
(recall that $P$ depends on $n$).  This is equal to the information
$I$, and is a unique measure of information\cite*{uniqueinfo}.
One can generalize the picture, by replacing the projectors with 
positive operators \cite*{Rains2001}. One is then interested in 
the positive operator $A$  of mimimal trace, satisfying 
\be
\trace(\rhon A)\geq 1-\epsilon
\label{eq:highfidpos}
\ee
It turns out that the rate is the same as in the case of projectors. 
Thus one can express the rate of concentration of information 
as follows 
\be
R^{conc} =\log d  - \lim_n {1\over n} \log \trace A\rhon
\ee
where $A$ is chosen to have minimal trace and satisfy the condition 
(\ref{eq:highfidpos}). In the following section we will see that 
the concentration of information in the distnt lab paradigm is 
connected with a similar quesiton,
however the operator $A$ will be suitably constrained.

\section{NLOCC maps and dual maps}

In entanglement theory, the paradigm is based on considering
Local Operations and Classical Communciation (LOCC maps).
Here we will define a map which we call NLOCC maps\cite*{nlocc}
for Noisy Local Operations and Classical
Communication.   
The motivation for using NLOCC maps, as opposed to the usual
LOCC maps is that here we are interested not in distilling singlets,
but in distilling pure product states, and so, care must be taken to
properly account for all pure states, including local ancillas, 
involved in any transformation. We will show that the dual 
map of an NLOCC map is (up to a factor) also an NLOCC map. 
We will need this  to represent distillation to product states
as compression with restricted means. 

\subsection{NLOCC maps}
NLOCC maps are any maps that 
can be composed  of the following maps:
\bee
\item local unitary transformations
\item adding quantum system in a {\it maximally mixed} state
\item discarding local subsystems (local partial trace)
\item sending subsystems down  completely decohering 
(dephasing) channels
\label{item:cl-channel}
\eee
The latter channel is of the form
\be
\varrho_{in}\to \varrho_{out}=\sum_iP_i\varrho_{in} P_i
\ee
where $P_i$ are one-dimensional projectors. 
For a qubit system, it acts as 
\be
\varrho_{in}=\left[\bea{cc}
\varrho_{11} & \varrho_{12} \cr
\varrho_{21} & \varrho_{22} \cr
\eea\right]
\to 
\varrho_{out}=\left[\bea{cc}
\varrho_{11} & 0 \\
0 & \varrho_{22}\\
\eea\right]
\ee
i.e. the state becomes classical, as if it has been measured.
Here, $\varrho_{in}$ is at the sender's site, while $\varrho_{out}$
is at the receiver's site. The operation (\ref{item:cl-channel}) can be 
disassembled into two parts: (i) local dephasing  
(at say, the sender's site) and (ii) sending qubits intact 
(through a noiseless quantum channel) to the receiver.
Thus suppose that  Alice and Bob share a  state 
$\varrho_{AB}\equiv \varrho_{A'A''B}$, and Alice decides to 
send subsystem $A''$ to Bob, down the dephasing channel.
The following action will have the same effect: Alice dephases locally the 
subsystem  $A''$
\be
\varrho_{A'A''B} \to \sum_i (P_i^{A''}\otimes I_{A'B} )
\,\varrho_{A'A''B}\,
(P_i^{A''}\otimes I_{A'B})
\ee
The state is now of the form 
\be
\varrho_{A'A''B}^{out}= \sum_ip_i P_i^{A''} \otimes \varrho^{A'B}_i
\ee
Thus the part $A''$ is classically correlated with the rest of 
the system (this is a stronger statement than to say that the 
state is separable  with respect to $A'':A'B$). 
Now Alice sends the system $A''$ 
to Bob through an ideal channel. Thus the final state differs 
from the state $\varrho_{A'A''B}^{out}$ only in that the 
system $A''$  is at Bob site.  It follows that the operation 
\ref{item:cl-channel} can be replaced by the following two operations
\bee
\item[4a)] Local dephasing
\item[4a)] Sending completely dephased subsystem. 
\eee

Note that the definition of NLOCC maps differ from LOCC only in that 
one cannot add local ancilla in any state, but only in maximally 
mixed one. One allows POVM's which involves adding a pure state
ancilla, but we do so by including the ancilla as part
of the initial state $\varrho$.   This difference causes us to 
consider in more detail the classical communication.  Indeed, 
since information becomes the resource, we have to take care 
also about the 
bits that carry information between the labs. The information  
represented  by  the carriers must be counted. Otherwise, given the 
initial state, we could supply an infinite amount of local information, 
just by multiplying the number of bits comunicated.  

The reader may have the impression that unlike in LOCC, we have disallowed 
making measurement, and that this is another big difference 
between our maps and LOCC. As a matter of fact, we do allow measurements,
however we care not only about the system to be measured but also
the measuring apparatus. This is 
natural approach when one counts pure states as a resource,
since a measuring device must initially be in a pure state to
be effective, and one must therefore take care to properly
account for this.  This is also natural from the thermodynamical
point of view: Maxwell's demon
was exorcised just by including itself into the description 
\cite*{Bennett82}.
%.,Plenio-Vitali}. 
%

\subsection{Dual maps}
Let us now describe the dual maps to the elementary NLOCC maps of the 
previous subsection, and show that they are also NLOCC maps
up to a factor. Consider a map 
\be
\Lambda:\bcal(\hcal_1)\to \bcal(\hcal_2)
\ee
The dual map
\be
\Lambda^\dagger:\bcal(\hcal_2)\to \bcal(\hcal_1)
\ee
is defined with repsect to the Hilbert-Schmidt scalar product
\be
\<A|B\> =\trace (A^\dagger B)
\ee
by the following relation
\be
\trace (A^\dagger \Lambda (B))=\trace (\Lambda^\dagger(A^\dagger) B)
\ee
for any operators $A\in \bcal(\hcal_1)$, $B\in\bcal(\hcal_2)$. 
If the map is completely positive i.e. it is of the form 
\be
\Lambda(\varrho)=\sum_iV_i\varrho V_i^\dagger
\label{eq:map}
\ee
its dual is of the form
\be
\Lambda^\dagger(\varrho)=\sum_iV_i^\dagger\varrho V_i
\label{eq:dual}
\ee
i.e. it is still completely positive, though perhaps not trace preserving. 
If the output dimensions are equal to each other, then 
the dual is trace preserving iff the map preserves the identity 
(is bistochastic). 
Out of our elementary maps, the maps with equal in and out dimensions are 
(1),(4a) and (4b). They also preserve the identity, hence their duals are 
valid physical operations.
Specifically,  the dual to a local unitary is again a local unitary 
(the inverse one), and the local dephasing  is self-dual,
i.e. is equal to its dual.  This can be seen simply by inspecting
(\ref{eq:map}) and (\ref{eq:dual}).
 
A dual to the map (4b)  is simply sending a qubit 
in the converse direction. Consider now the maps (2) and (3). They 
are not trace preserving, but the only reason  is that the in and out 
dimensions are not equal, hence, as trace preserving maps, 
they cannot preserve identity. However our maps (2) and (3) 
preserve the maximally  mixed state. As a result the duals will
be trace preserving up to a factor. For a state $\varrho$ we have 
\ben
&&\trace(\Lambda^\dagger(\varrho))=
\trace(\Lambda^\dagger(\varrho)I_{out})=
 \trace(\varrho\Lambda(I_{out}))= \\ \nonumber
&& =d_{out}\trace(\varrho\Lambda(I_{out}/d_{out})=
d_{out}\trace(\varrho I_{in}/d_{in})= {d_{out}\over d_{in}}
\label{eq:trace-pmm}
\een
where $d_{in}$ and $d_{out}$ are the dimensions of input and output 
Hilbert  spaces $\hcal_1$ and $\hcal_2$, respectively. Consequently
the map $\Gamma={d_{in}\over d_{out}}\Lambda^\dagger$ 
is a valid physical operation. 

One then easily finds that up to a factor the map (2) of
adding a maximally mixed state is dual to map (3) of local
partial trace 
and {\it vice versa}. More precisely if 
\be
\Lambda(\varrho_{A'A''B}) =\varrho_{A'B}
\ee
is (local) partial trace then the dual map is given by 
\be
\Lambda^\dagger(\varrho_{A'B}) =\varrho_{A'B}\otimes I_{A''} =
\dim_{A''}\biggl[\varrho_{A'B}\otimes {I_{A''}\over d_{A''}}\biggr]
\ee

\section{Distilling pure product states and compression}

In this section we will represent distillation of product states 
by NLOCC  as compression with restricted means. 
First of all it is clear that in comparison with the usual compression
situation described 
in section \ref{sec:single-system}, we have a restricted class of 
operations. As we shall see, this will lead us to optimize 
over suitably constrained positive operators $A$. We will follow 
Rains approach \cite*{Rains2001} (see also \cite*{WernerPPT}), 
similarly as it was applied  in  Ref. \cite*{nlocc}.
Our goal is the following: to obtain  the maximal number of 
product states out of $\rhon$, where $\varrho_{AB}$ 
is a state on $C^d\otimes C^d$. I.e. one wants to 
maximize the rate $r=m/n$ in the transition
\be
\rhonab \to \prodm
\ee
I.e., instead of distilling singlets, we wish to distill
local pure states.
Here $P_{|00\>}=|00\>\<00|$ is a state of two qubits.
Consider a given  protocol of concentration of information,
i.e. the map $\Lambda$ that takes $\rhonab$ to some state 
close to the required $\prodm$.  
(More precisely it is family of maps indexed by $n$).
We impose some fixed rate of transition $r=m/n$, 
and require the fidelity of transition to tend asymptotically (for 
$n\to\infty)$)  to 1
\be
F=\trace \bigl[\prodm\Lambda(\rhonab)\bigr]\geq 1-\epsilon.
\label{eq:fid-nlocc}
\ee
The condition for fidelity will give us a restriction on possible rates $r$.
We rewrite the fidelity as 
\be
F=\trace \bigl[\Lambda^\dagger(\prodm)\rhonab\bigr].
\ee
Consider now the operator $\Pi=\Lambda^\dagger(\prodm)$.
Let us evaluate its trace. According to formula (\ref{eq:trace-pmm})
we obtain that 
\be
\trace \Pi={d_{in}\over d_{out}} = {d^{2n}\over 2^m}
\ee
Hence the asymptotic rate of information concentration is given 
by 
\be
r=2\log d - {1\over n}\log \trace[\Pi \rhonab]
\ee 
The maximal possible rate is obtained if in the formula above 
we put optimal operators $\Pi$, that
(i)  satisfy high  fidelity condition  (\ref{eq:fid-nlocc}), and  
(ii) can be obtained by action of $\Lambda^\dagger$ from 
$\prodm$, where $\Lambda$ is NLOCC.  

We can now interprete the action  of concentration of information to 
local form as follows. Alice and Bob, want  to 
separate noise from purity. Normally they should perform
the measurement $\{P,I-P\}$ to project the state onto the typical subspace.
However the projector $P$ onto the typical subspace, or 
the projector $I-P$ may be entangled. 
Also the state $\psi$ of decomposition 
into $\hcal_s$ and $\hcal_r$ can be entangled. 
Then in general they cannot implement such a measurement. 
%There may be two reasons why they cannot do this. 

Indeed, Alice and Bob can make only 
such measurements  that are allowed by LOCC, and these
measurements may not commute with the projector $P$. 
Thus, to be able to compress the state, Alice and Bob have to destroy it 
making it "classical". Then using classical communication they can
transform  it in a lossless way to product form, and perform local 
compression.

Note that what makes this compression less efficient 
is the fact
that under LOCC the implementable measurements may not commute with
the projector $P$.  It is quite possible,
that although $P$ is not implementable, one can
%however another situation that they are able 
make a {\it finer measurement}
that will commute with the above $P$. This is okay in this
paradigm, since the pure states remain pure (the signal is
damaged, but we do not care about that in this ``objective''
compression scheme).
If the measurement is finer only in the part $P$, so that it is of the form
$\{P_1,\ldots,P_n, I-P\}$, then the concentration is done as well 
by NLOCC as by means of NO: the pure state is untouched, even though the 
signal state
is cut into incoherent pieces. However if it is finer in the second part $I-P$, 
then this may damage the state $\psi$. 
%Note however, that there is an important difference between a measurement
%using the highly degenerate projector $P$, and some finer measurement.  
%The difference is not noticable because we use NLOCC paradigm
%In the former case, the measuring device, which could be taken in a pure
%state
%We will come back to this
%point in a moment.

%First, 
Let us now go back to discussion of the constraints (i) and (ii) above
that $\Pi$ must satisfy.
 The first one is 
the same as the one of eq. (\ref{eq:highfidpos}) in the usual compression 
scheme, with positive operators instead of projectors. Thus 
we should consider the second constraint that was not present in 
the usual compression case. First of all, the operator $\Pi$ must be,
up to a factor, a separable state. Thus it must be a mixture of 
product positive operators. This is because the map $\Lambda^\dagger$ 
is up to a factor, an NLOCC operation, hence it can  not transform 
a product  state $\prodm$ into an entangled state. 

This suggests considering the maximal rate possible with operators $\Pi$ 
being separable ones. 
%\begin{definition}
%{\tt Jakos to nazwac}
%\end{definition}
This is certainly an upper bound for the optimal rate of 
concentration of information. It closely resembles compression 
of an unknown source which we now describe roughly \cite*{CoverThomas}. Suppose we 
have a source $\rhon$
but think it is $\sigman$. We then take projectors $P$,
which are suitable for $\sigma$. Of course it will not work for our source.
However, we can try to take projectors of greater dimension 
built in an analogous way to $P$. The question is how large should  
the subspace be, to be good also for the state $\rhon$?
The answer is that the compression rate will now be worse.
The penalty will be $S(\varrho|\sigma)=\trace \varrho\log \varrho-\trace 
\varrho\log\sigma$. Thus the rate of "objective" information 
concentration will be $\log d -S(\varrho) - S(\varrho|\sigma)$.
We thus get the bound for the rate under NLOCC
\be
r \leq \log d -S(\varrho) -\inf_{\sigma\in sep} S(\varrho|\sigma)
\label{eq:ubound}
\ee

with $\sigma$ taken from the set of seperable states.
In view of the discussion above, this scheme of concentration 
of information to a local form is a sort of compression 
while imagining that the state is separable, or at least 
while using "separable tools".

If such a bound were achievable, then we would essentially
have that the penalty one pays under NLOCC for compression,
is the relative entropy of entanglement\cite*{PlenioVedral1998}.
However, it seems that instead, the penalty is given not only
by the entanglement, but also, by so called non-locality without entanglement
\cite*{Bennett-nlwe}.
%However, the bound Eq. (\ref{eq:ubound}), does not appear
%to be achievable, because it appears that the concentration by means 
%of separable $\Pi$ is better than the actual concentration 
%by NLOCC maps.  
Whether this is the case,  
amounts to the question:  can we produce by use of the map 
$\Gamma={d_{out}\over {d_in}}\Lambda^\dagger$,   where $\Lambda$ is NLOCC,
all possible separable  states of the input system  out of $\prodm$?
The answer seem to be negative: Namely there are additional 
constraints. Since the map $\Gamma$ is NLOCC, it cannot increase the 
amount of information. The initial information is $2m$ bits (this is 
actually the final information in the real scenario, yet here we 
deal with the dual scenario). {\it Thus one can obtain only such 
separable states, that 
can be prepared by NLOCC from 2m bits of information.}
The question of {\it information of formation} was considered in Ref. 
\cite*{phase}. It was pointed out there, that 
the local information needed to create a separable state 
may be greater than $n-S$. This is because, a separable state 
may be a mixture of product states, that are locally non-orthogonal. 
Then to create such states,  one will be forced to use some 
irreversible operations. In particular, one expects
density matrices whose eigenbasis are the ``sausage states'' of
Ref. \cite*{Bennett-nlwe} to not be preparable with only
$n-S$ bits.  
%In other words, the 
%penalty of the rate of compression under NLOCC feels not only
%entanglement, but also, various types of non-locality without
%entanglement.
Thus the fact that we deal with NLOCC operations rather than LOCC 
ones enter the scene twice. First, it restricts the trace of 
operator $\Pi$ making the problem nontrivial. Second, it 
restricts the class of separable states, that corresponds to $\Pi$
up to a factor.

\section{Conclusion}

By applying Rains method, we have shown how to obtain
bounds on the rate of compression of information when
in the distant labs paradigm.
%To obtain the rate of compression, one should therefore not choose 
%$\sigma$ from the set of seperable states, but instead choose
%$\sigma$ from some smaller set implementable by NLOCC.  
In \cite*{nlocc}
it was conjectured that the maximum compression rate under NLOCC was
\be
r = \log d -S(\varrho) -\inf_{\sigma\in IPB} S(\varrho|\sigma)
%$\label{eq:ubound}
\ee
where IPB is the set of states whose eigenbasis were orthogonal projectors
implementable under NLOCC.  i.e. projectors which can be achieved from 
the standard projector $\prodm$.
It would be very valuable to prove such a conjecture.

Here, we have explored an ``objective'' scheme of compression of information,
where instead of protecting a signal, one instead wishes to obtain
pure states. 
It also may be interesting to explore the ``subjective'' scheme of compression
in the distant labs paradigm.  Such scenarious might lead one to consider
various quantum versions of the Slepian-Wolf theorem\cite*{slepian-wolf} (c.f.
 \cite*{Winter1999PhD,igor2002,keiji2002})

In Slepian-Wolf encoding, two people in distant labs attempt to compress a
signal.  They then meet at some later time to decompress it.  In a
quantum version of such a scenario, the compression rate would
depend heavily on the ensemble.  If the signal states were entangled, then
very little compression could take place, but if the signal states were
classical states, then the entire state should be compressable.  The rate
of compression would therefore not be a universal function of the density
matrix, but would instead depend on the ensemble. Further exploration of a 
quantum Slepian-Wolf theorem may be of interest.

\begin{acknowledgments}
This work is supported by EU grants RESQ, contract No. IST-2001-37559,
and QUPRODIS, contract No. IST-2001-38877.
J.O. acknowledges the support of
the Lady Davis Fellowship Trust, and
grant No. 129/00-1 of the Israel Science Foundation.
\end{acknowledgments}
\bibliography{d:/prace/referencje/refmich}%,c:/prace/referencje/refjono}
\end{document}